# Perturbative approach to the self-focusing of intense X-ray laser beam propagating in thermal quantum plasma

*R. Roozehdar Mogaddam[1], N. Sepehri Javan[2,\*], K. Javidan[1,†], H. Mohammadzadeh[2]*

[1]*Department of Physics, Ferdowsi University of Mashhad, 91775-1436 Mashhad, Iran*

[2]*Department of Physics, University of Mohaghegh Ardabili, PO Box 179, Ardabil, Iran*

*Corresponding Authors:\*sepehri_javan@uma.ac.ir, † javidan@um.ac.ir*

**Abstract**

In this theoretical study, the problem of self-focusing of an X-ray intense laser beam in the thermal quantum plasma is studied. Using a relativistic fluid model and taking into account the hydrodynamic pressure of degenerate electrons in the zero temperature limit, the nonlinear momentum equation of electrons is solved by means of a perturbative method and the nonlinear current density of the relativistic degenerate electrons is obtained. Saving only the third-order nonlinearity of the laser beam amplitude, a nonlinear equation describing the interaction of laser beam with the quantum plasma is derived. It is shown that considering the nonlinearity of system through solving nonlinear equation of degenerate electron leads to the originally different wave equation in comparison with outcomes of the approach in which the permittivity of longitudinal waves of quantum plasma is problematically extended to the relativistic case. The evolution of laser beam spot size with Gaussian profile is considered and the effect of quantum terms on the self-focusing quality is studied. It is shown that considering quantum effects leads to the decrease in the self-focusing property and the effect of Bohm tunneling potential is more dominant than the degenerate electrons pressure term.

## I. Introduction

After achieving high power lasers in the recent decades, theoretical and experimental studies of laser plasma interaction have become more attractive, because of its important technological applications, including charged particles acceleration in plasma wake-field [1-3], laser fusion [4-6], higher harmonics generation [7-11], magnetic field generation [12], X-ray lasers [13-15], etc. Plasma constituents receive considerable amounts of energy during their interactions with laser beam and find relativistic velocities. This process substantially modifies the physical properties of the medium, which causes fundamental changes in the propagation characteristics of the laser beam. Deformation of longitudinal or perpendicular distribution of laser beam intensity during its interaction with plasma can be studied in the context of some nonlinear phenomena such as modulation [16, 17], and filamentaion [18, 19] instabilities and/or self-focusing (SF) as well. The SF is a well-known nonlinear optical phenomenon which occurs due to the modification of the refractive index due to the interaction with a strong electromagnetic radiation [20-22]. In high intensities of laser beam, the refractive index of plasma can be considerably increased by the motion of electrons interacting with the electric field of the laser beam. On the other hand, at the presence of an electromagnetic wave having an initial transverse intensity profile, formation of variable refractive index can localize the beam like a positive focusing lens. Recent advances in laser technology, have enabled the observation of SF phenomenon within the interaction of intense laser pulses with plasmas. There have been presented a lot of theoretical and experimental works dealing with the SF of laser beam in the classical plasma and pair plasma as well [23-41]. A review of related literature shows that quantum plasmas are encountered in many environments, for example, in very small-sized electronic devices [42], carbon nanotubes [43], quantum dots [44], quantum well and quantum diodes [45, 46], biophotonics [47], dense astrophysical environments [48] and in the interaction of laser with solid targets [49]. It may be noted that, there are a few studies related to the SF in quantum plasmas. For the first time, the SF of a Gaussian laser beam in the relativistic cold quantum plasma is theoretically studied by Patil et al. in 2013 [50]. The study was started by introducing the dielectric constant for cold quantum plasma derived by Jung and Murakami [51] and using it in the nonlinear wave equation. Using the same method, thermal effect caused by the zero temperature limit pressure of electrons on the SF in quantum plasma was investigated in the same year as well [52]. Zare et al. [53] considered effects of electro-ion collision on the SF of warm quantum plasma, following the same approach of Refs. [50] and [52]. Extension of SF problem in the cold and warm quantum plasma by considering a nonhomogeneous medium was accomplished in some theoretical investigations [54-56]. Effects of the magnetization of quantum plasma on the SF have been recently studied by Aggarwal et al [57] in the weakly relativistic regime of laser beam intensity.

All above mentioned theoretical studies on the SF of quantum plasma are founded on a same formalism in which the linear dielectric constant of quantum plasma is extended to the physical circumstances of a nonlinear problem for the construction of Kerr nonlinearity. In all cases, they have used the dielectric constant of plasma related to the longitudinal oscillations of quantum plasma which is valid for the space-charge waves. In contradiction with classical plasma, the electric permittivity of the quantum plasma medium for the longitudinal and transverse waves are different [58]. Therefore, using the electric constant of longitudinal waves for the propagation of transverse electromagnetic waves can be problematic as we will deal it in detail in the section IV. In this theoretical work, for the first time, we investigate the SF problem of quantum plasma by considering dynamics of electrons under interaction with electromagnetic fields of the laser beam. Nonlinear current density of electrons is derived and a nonlinear third-order equation describing the evolution of laser beam amplitude is obtained. It is shown that independently deriving nonlinearity via considering degenerate relativistic electrons dynamics and neither through extension of electric constant of longitudinal waves, can basically lead to the different results. For example, our findings show that considering quantum effects leads to a decrease in the nonlinearity of plasma and consequently decrease in its focusing quality which is in contradiction with the results of previous studies (for example, see Ref. [50]).

**II. Deriving nonlinear wave equation**

Let us consider the propagation of linearly polarized laser beam in the quantum plasma. We take the electric field of laser beam along the $x$-axis as following

$$\mathbf{E} \equiv \hat{E}\hat{\mathbf{e}}_{\mathbf{x}} \cos(k_0 z - \omega_0 t), \tag{1}$$

where $\hat{E}$, $\hat{\mathbf{e}}_{\mathbf{x}}$, $\omega_0$ and $k_0$ are the slowly varying amplitude, the unit vector of x-axis, frequency and wave number of the laser beam, respectively. The slow variation of $\hat{E}$ means that $\left|\frac{1}{\omega}\frac{\partial \hat{E}}{\partial t}\right| << \left|\hat{E}\right|$, $\left|\frac{1}{k}\frac{\partial \hat{E}}{\partial z}\right| << \left|\hat{E}\right|$. According to the Maxwell's equations for a linearly-polarized electromagnetic wave, the electric and magnetic fields of laser beam and wavenumber vector are perpendicular to each other and they constitute a right-handed system. From Faraday's equation, i.e. $\nabla \times \mathbf{E} = -(1/c)\partial \mathbf{B}/\partial t$, the magnetic field of the laser beam can be obtained as

$$\mathbf{B} \equiv \frac{k_0 c}{\omega_0} \hat{E}\hat{\mathbf{e}}_{\mathbf{y}} \cos(k_0 z - \omega_0 t), \tag{2}$$

where $c$ and $\hat{\mathbf{e}}_y$ are the speed of light and the unit vector of y-axis, respectively.

Some straightforward mathematical operations on Maxwell's equations lead to the following equation for the electric field of the laser beam

$$\left(\nabla^2 - \frac{1}{c^2}\frac{\partial^2}{\partial t^2}\right)\mathbf{E} = \frac{4\pi}{c^2}\frac{\partial \mathbf{J}}{\partial t}, \qquad (3)$$

where the electron current density is $\mathbf{J} = -en\mathbf{v}$, in which $n$ is the density of electrons, $\mathbf{v}$ is the electron velocity and $e$ is the magnitude of electron charge. For deriving current density and describing dynamics of electrons, we start with the following relativistic momentum and continuity equations for the fluid quantum plasma

$$\frac{\partial(\gamma\mathbf{v})}{\partial t} + \mathbf{v}.\nabla(\gamma\mathbf{v}) = -\frac{e}{m_0}\left(\mathbf{E} + \frac{\mathbf{v}}{c}\times\mathbf{B}\right) + \frac{\hbar^2}{2m_0^2}\nabla\left(\frac{\nabla^2\sqrt{n}}{\sqrt{n}}\right) - \frac{1}{m_0 n}\nabla\Pi, \qquad (4)$$

$$\frac{\partial n}{\partial t} + \nabla.(n\mathbf{v}) = 0, \qquad (5)$$

where $m_0$ and $\gamma = (1 - v^2/c^2)^{-1/2}$ are the electron rest mass and the relativistic Lorentz factor of electrons, respectively, while $h$ is the Planck's constant ($\hbar = h/2\pi$). The second term at the right-hand side of Eq. (4) is the Bohm potential term which is appeared because of tunneling effect in quantum plasmas [59]. The last term is the electron hydrodynamic pressure in the zero temperature limit, which can be obtained from the following equation [60]

$$\Pi = \frac{m_0 V_F^2 n_0}{3}\left(\frac{n}{n_0}\right)^{D+2/D}, \qquad (6)$$

where $V_F = (2\varepsilon_F/m_0)^{1/2}$ is the Fermi speed, $\varepsilon_F = \frac{\hbar^2}{2m_0}(3\pi^2)^{2/3}n^{2/3}$ is the Fermi energy of cold electron gas, $n_0$ is the unperturbed electron density and $D$ is the number of degrees of freedom which for the one-dimensional evolution of electron fluid is the unity. Here, we considered the zero temperature limit for the degenerated electron gas, whose equilibrium Wigner function is [59]

$$f_0(v) = \begin{cases} \dfrac{\Gamma(1+D/2)n_0}{\pi^{D/2}V_F^D}, & |v| < V_F \\ 0, & |v| > V_F \end{cases}, \tag{7}$$

where, $\Gamma(x)$ is the well-known gamma function. It is the simplest approach for a degenerate plasma where we consider a uniform electron distribution for kinetic energy smaller than the Fermi energy and there is no particle above the Fermi level.

For solving equations (4) and (5), we employ the well-known perturbative method. In this method we assume that each parameter is the summation of different orders with respect to the amplitude of laser fields as follows

$$\mathbf{v} = \mathbf{v}^{(0)}(=0) + \mathbf{v}^{(1)} + \mathbf{v}^{(2)} + ..., \tag{8-a}$$

$$n = n^{(0)}(=n_0) + n^{(1)} + n^{(2)} + ..., \tag{8-b}$$

$$\gamma = \gamma^{(0)}(=1) + \gamma^{(1)}(=0) + \gamma^{(2)} + ..., \tag{8-c}$$

$$\Pi = \Pi^{(0)} + \Pi^{(1)} + \Pi^{(2)} + .... \tag{8-d}$$

In this perturbative approach, the zeroth-order parameters which are indicated by the superscript (0) are unperturbed values of related parameters and specify the situation of the plasma system before interaction with the laser beam. The first order parameters which are indicated by the superscript (1) are caused by the terms proportional to the amplitude of the laser beam fields in the linear regime. The source of the first order parameters is the laser electric field force $-e\mathbf{E}$. The second order parameters which are related to the nonlinear terms, are proportional to the squared amplitude $E^2$. Their source is the magnetic Lorentz force term $-e\dfrac{\mathbf{v}^{(1)}}{c} \times \mathbf{B}^{(1)} \approx \mathbf{E} \times \mathbf{B} \approx E^2$. Sources of other higher-order parameters can be found by a similar method and one can deduce that generally, the nth-order parameters are proportional to $E^n$.

Substituting Eq. (8-b) in Eq. (6) and expanding it with respect to the small perturbative parameters lead to

$$\frac{1}{m_0 n}\nabla\Pi = \beta\frac{\partial n^{(1)}}{\partial z} - \frac{\beta n^{(1)}}{n_0}\frac{\partial n^{(1)}}{\partial z} + \beta\frac{\partial n^{(2)}}{\partial z} + ..., \tag{9}$$

where $\beta = 11(3\pi^2)^{2/3}\hbar^2/(9m_0^2 n_0^{1/3})$.

Using Eqs. (8-a)-(8-d) in Eq. (4) and separating first-order terms lead to the following first-order momentum equation for electrons

$$\frac{\partial v_x^{(1)}}{\partial t} = -\frac{e}{m_0}E^{(1)}, \tag{10}$$

where the first-order electric field $E^{(1)}$ is the laser electric field described by Eq. (1). The solution of Eq. (10) is

$$v_x^{(1)} = ca\sin(k_0 z - \omega_0 t), \tag{11}$$

where $a = e\hat{E}/(m_0 c\omega_0)$ is the normalized laser beam amplitude.

A similar procedure can be employed to obtain the second-order momentum equation as

$$\frac{\partial v_z^{(2)}}{\partial t} + v_z^{(1)}\frac{\partial v_z^{(1)}}{\partial z} = \frac{-e}{m_0 c}v_x^{(1)}B^{(1)} + \frac{\hbar^2}{4m_0^2 n_0}\left[\frac{\partial^3 n^{(2)}}{\partial z^3} - \frac{1}{2n_0}\frac{\partial}{\partial z}\left(n^{(1)}\frac{\partial^2 n^{(1)}}{\partial z^2}\right) - \frac{1}{4n_0}\frac{\partial^3 n^{(1)2}}{\partial z^3}\right] + \beta\left[\frac{n^{(1)}}{n_0}\frac{\partial n^{(1)}}{\partial z} - \frac{\partial n^{(2)}}{\partial z}\right]$$
.(12)

As $\nabla . E^{(1)} = 0$ then $n^{(1)}$ is zero, there is no first-order perturbation of the electron density and according to Eq. (5) the first-order longitudinal velocity $v_z^{(1)}$ is zero as well. Second-order longitudinal displacements can produce electron density modulation which one can obtain the following equation by expanding the continuity Eq. (5)

$$\frac{\partial n^{(2)}}{\partial t} + n_0 \frac{\partial v_z^{(2)}}{\partial z} = 0. \tag{13}$$

Simultaneous solutions of Eqs. (12) and (13) can be easily obtained as

$$v_z^{(2)} = \frac{c^2 a^2 \omega_0 k_0}{-4\omega_0^2 + \frac{4\hbar^2 k_0^4}{m_0^2} + 4n^{(0)}\beta k_0^2}\cos(2k_0 z - 2\omega_0 t), \tag{14}$$

$$n^{(2)} = \frac{n^{(0)} c^2 a^2 k_0^2}{-4\omega_0^2 + \frac{4\hbar^2 k_0^4}{m^2} + 4n^{(0)}\beta k_0^2} \cos(2k_0 z - 2\omega_0 t). \tag{15}$$

The third-order momentum equation becomes

$$\frac{\partial v_x^{(3)}}{\partial t} + v_z^{(1)} \frac{\partial v_x^{(2)}}{\partial z} + v_z^{(2)} \frac{\partial v_x^{(1)}}{\partial z} + \frac{1}{2} \frac{\partial}{\partial t}\left(\frac{v_x^{(1)3}}{c^2}\right) = \frac{e}{mc}\left(v_z^{(2)} B^{(1)}\right), \tag{16}$$

where the first and forth terms of the left-hand side of Eq. (6) come from the relativistic terms $\frac{\partial}{\partial t}(\gamma^{(0)} v_x^{(3)}) = \frac{\partial v_x^{(3)}}{\partial t}$ and $\frac{\partial}{\partial t}(\gamma^{(2)} v_x^{(1)}) = \frac{1}{2}\frac{\partial}{\partial t}\left(\frac{v_x^{(1)3}}{c^2}\right)$ while $B^{(1)}$ is the first-order magnetic field of the laser beam described by Eq. (2). The solution of Eq. (16) for the third-order transverse velocity is the following

$$v_x^{(3)} = -\frac{3}{8} ca^3 \sin(k_0 z - \omega_0 t) + \frac{1}{8} ca^3 \sin(3k_0 z - 3\omega_0 t), \tag{17}$$

which includes both fundamental and third harmonics. Substituting Eqs. (11), (15) and (17) in the nonlinear current density $j_x = -e(n^{(0)} v_x^{(1)} + n^{(0)} v_x^{(3)} + n^{(2)} v_x^{(1)})$ and separating only the fundamental harmonics give

$$j_x = -ecn_0 a\left[1 - \frac{|a|^2}{8}\left(3 - \frac{c^2 k_0^2 / \omega_0^2}{1 - \frac{\hbar^2 k_0^4}{m^2 \omega_0^2} - \frac{n_0 \beta k_0^2}{\omega_0^2}}\right)\right] \sin(k_0 z - \omega_0 t). \tag{18}$$

Nonlinear current density of Eq. (18) which introduces the Kerr effect, is the source of several nonlinear phenomena including SF. Ignoring the second and third terms in the denominator of the nonlinear term in Eq. (18) leads to the relativistic nonlinearity of the ordinary classical plasma which is in accordance with the previous studies [33, 34]. The term $\hbar^2 k_0^4 /(m^2 \omega_0^2)$ comes from the Bohm potential term while the term $n_0 \beta k_0^2 / \omega_0^2$ is related to hydrodynamic pressure of degenerate electrons.

Substituting nonlinear current density from Eq. (18) in the wave equation (3) yields the following equation for the nonlinear dynamics of laser beam amplitude

$$\left(\nabla^2 - \frac{1}{c^2}\frac{\partial^2}{\partial t^2}\right) a\cos(k_0 z - \omega_0 t) = k_p^2 \left(1 - |a|^2 N\right) a\cos(k_0 z - \omega_0 t), \quad (19)$$

where $k_p = \omega_p / c$, $\omega_p = \sqrt{4\pi n_0 e^2 / m_0}$ is the plasma frequency and

$$N = \frac{3}{8} - \frac{c^2 k_0^2 / \omega_0^2}{8\left(1 - \frac{\hbar^2 k_0^4}{m^2 \omega_0^2} - \frac{n_0 \beta k_0^2}{\omega_0^2}\right)}. \quad (20)$$

In the non-interactional regime of laser beam when we neglect variations of the laser beam amplitude and take it constant, after taking temporal and spatial derivatives of cosines function in Eq. (19), one can write the following nonlinear dispersion relation for the quantum plasma

$$D_{NL} = k_0^2 - \frac{\omega_0^2}{c^2} + \frac{\omega_p^2}{c^2} - \frac{\omega_p^2}{c^2}\left[\frac{3}{8} - \frac{c^2 k_0^2 / \omega_0^2}{8\left(1 - \frac{\hbar^2 k_0^4}{m^2 \omega_0^2} - \frac{n_0 \beta k_0^2}{\omega_0^2}\right)}\right]|a|^2 = 0. \quad (21)$$

In the linear limit, where $|a|^2 \to 0$, Eq. (20) reduces to the well-known dispersion relation

$$k_0 = \frac{\omega_0}{c}\left(1 - \frac{\omega_P^2}{\omega_0^2}\right)^{\frac{1}{2}}. \quad (22)$$

As it is clear from Eq. (22), linear dispersion relation of a quantum plasma for the transverse electromagnetic waves is the same of the classical one. As indicated in the Ref. [28], one can find out that, at the presence of electromagnetic waves in a non-magnetized quantum plasma, there is no contribution of quantum effects on dispersion in the linear limit, while for the longitudinal space-charge waves, quantum effects appear even in the linear limit.

**III. Envelope evolution**

By considering the nonlinear wave equation (19) which contains the third-order Kerr nonlinearity, a lot of nonlinear phenomena including SF can be studied. There is a well-known so-called Source Dependent Expansion (SDE) method [61] for describing the stationary evolution of the laser beam envelope under SF mechanism. According to this method the laser amplitude is expanded as a series of Laguerre-Gaussian source-dependent modes as $a(r,z) = \sum_m \hat{a}_m L_m(r^2/r_s^2)\exp[-(1-i\alpha_s)r^2/r_s^2]$, in which $\hat{a}_m(z)$ is a complex amplitude, $r_s(z)$ is the spot size, $\alpha_s(z) = k_0 r_s^2 / 2R_c$ is related to the curvature $R_c$ associated with the wave-front and $L_m(r^2/r_s^2)$ is a Laguerre polynomial of order $m$. Considering only the lowest order Gaussian mode ($m=0$) in the summation leads to the following simple Gaussian form for the laser beam amplitude

$$a(r,z) = a_0 \exp\left(-i\alpha_s \frac{r^2}{r_s^2} + i\theta_0\right)\exp\left(-\frac{r^2}{r_s^2}\right), \tag{23}$$

where $\theta_0$ is the phase of the zero-order complex amplitude and $a_0$ is its real amplitude. Focusing or defocusing of laser beam is determined by the spatial evolution of the parameter $r_s(z)$. Decrease in the spot size parameter $r_s(z)$ in terms of the increase in the $z$ represents the focusing of laser beam during its propagation. The first exponential term in Eq. (23) is a phase term which has no role in the laser beam intensity distribution, therefore ignoring the procedure of finding $\theta_0$, we focus on finding the spot size parameter $r_s(z)$. In the paraxial approximation when $(r/r_s)^2 < 1$, substituting Eq. (23) in the wave equation (19) yields the following differential equation for the evolution of the laser spot size [61]

$$\frac{\partial^2 r_s}{\partial z^2} = \frac{4}{k_0^2 r_s^3}\left(1 - \frac{k_p^2 a_0^2 r_0^2}{8} N\right). \tag{24}$$

where $r_s(z=0) = r_0$ is the initial laser beam spot size. The first term on the right-hand side of Eq. (24) is originated from the vacuum diffraction while the source of the second term is the Kerr nonlinear effect which is related to the different factors such as the density perturbation, relativistic mass variation and quantum effects as well. The normalized power of the laser beam is introduced as $P/P_c = (1/8)k_p^2 a_0^2 r_0^2 N$ which under satisfying the condition $P > P_c$ the focusing characteristic related to the nonlinearity of the medium overcomes the diffraction effect and the SF occurs. In fact, the term $p_c = 2\pi^2 c^5 m_0^2 /(k_p^2 \lambda^2 e^2 N)$ (where $\lambda$ is the wave-length of the laser beam) is the critical power for starting the SF of electromagnetic waves in the quantum plasma.

Straightforward mathematical operations reveal that the solution of Eq. (23) is

$$\frac{r_s^2}{r_0^2} = 1 + \left(1 - \frac{P}{P_c}\right)\frac{z^2}{Z_R^2}, \tag{25}$$

where $Z_R = k_0 r_0^2 /2$ is the Rayleigh length which is the well-known characteristic length for the wave diffraction. It is better to mention that the procedure of obtaining Eqs. (24) and (25) are similar with those of our previous studies [35-37] and in order to not to confuse readers, it has been repeated again.

## IV. Some points about the electric permittivity of quantum plasma

Let us start with the linear dispersion equation of longitudinal and transverse waves propagating through the unmagnetized thermal quantum plasma. For the longitudinal oscillations of space-charge or Langmuir wave, one can obtain the following equation using the Wigner-Poisson equations [62]

$$\varepsilon_\parallel = 1 - \frac{\omega_p^2}{\omega^2 - k^2 V_F^2 - \alpha k^4} = 0, \tag{26}$$

where $\alpha = \hbar^2 / 4m_0^2$. In the limit of non-thermal quantum plasma, Eq. (26) confirms the results of Ref. [54]

$$\varepsilon_\parallel = 1 - \frac{\omega_p^2}{\omega^2} \frac{1}{(1 - \lambda_q^4 k^4 \omega_p^2 / \omega^2)} = 0, \tag{27}$$

which is presented in different form and $\lambda_q = (\hbar^2/4m_0^2\omega_p^2)^{1/4}$ is the quantum wavelength of electron. With neglecting quantum effects in Eq. (25), viz. $\alpha, V_F \to 0$ or $\hbar \to 0$, the well-known classical dispersion relation is achieved

$$1 - \frac{\omega_p^2}{\omega^2} = 0. \tag{28}$$

For transverse electromagnetic quantum plasma waves in the linear limit, dispersion relation keeps its classical form and there is no quantum effects contribution. The form of the linear dispersion relation for the circularly polarized electromagnetic waves in the magnetized quantum plasma for the parallel propagation even is the same of classical case, however for the perpendicular propagation it differs from the classical one [58]. Using our Eq. (21) obtained for the quantum plasma with hydrodynamic electron pressure, one can obtain the following expression for the nonlinear refractive index for the transverse electromagnetic waves

$$\varepsilon_{NL} = n_{NL}^2 = \frac{k_0^2 c^2}{\omega_0^2} = 1 - \frac{\omega_p^2}{\omega_0^2} + \frac{\omega_p^2}{\omega_0^2}\left[\frac{3}{8} - \frac{c^2 k_0^2/\omega_0^2}{8\left(1 - \frac{\hbar^2 k_0^4}{m^2\omega_0^2} - \frac{n_0 \beta k_0^2}{\omega_0^2}\right)}\right]|a|^2, \tag{29}$$

which after using well-known definition $\varepsilon_{NL} = \varepsilon_0 + \Phi(aa^*)$ for the nonlinear refractive index, it can be presented as

$$\varepsilon_0 = 1 - \frac{\omega_p^2}{\omega_0^2}, \quad \Phi = \frac{\omega_p^2}{\omega_0^2}\left[\frac{3}{8} - \frac{c^2 k_0^2/\omega_0^2}{8\left(1 - \frac{\hbar^2 k_0^4}{m^2\omega_0^2} - \frac{n_0 \beta k_0^2}{\omega_0^2}\right)}\right]|a|^2. \tag{30}$$

The nonlinear permittivity $\varepsilon_{NL}$ is the fundamental physical parameter for analyzing the SF in the common approaches, one of which was used in the previous section. Here, we obtained this parameter by considering dynamics of electrons under interaction with laser beam fields and each other via hydrodynamic pressure of degenerate electrons as well. As we said earlier, all the mentioned theoretical works related to the SF in quantum plasma have employed dielectric function of longitudinal potential waves of quantum plasma extending it to the relativistic regime. For example, Patil et al. [46] have extended dielectric function of Eq. (27) to the relativistic regime by replacing $\omega_p^2$ with $\omega_p^2/\gamma$, where $\gamma = \sqrt{1+|a|^2}$ is the Lorentz relativistic factor, and used the following expression for the permittivity of transverse electromagnetic wave

$$\varepsilon = 1 - \frac{\omega_p^2}{\gamma \omega^2} \frac{1}{1-\delta/\gamma}, \tag{31}$$

where $\delta = \pi^2 \hbar^2 k_0^4 / m_0^2 \omega^2$. In Eq. (27), expanding $\gamma$ with respect to the small parameter of normalized laser beam intensity $|a|^2$ leads to

$$\varepsilon_0 = 1 - \frac{\omega_p^2}{\omega_0^2}(1+\delta), \quad \Phi = \frac{\omega_p^2}{\omega_0^2}\left(\frac{1}{2}+\delta\right)|a|^2, \tag{32}$$

which does not confirm our results of Eq. (30) in the limit of zero electron pressure $\beta \to 0$. In the limit $\hbar \to 0$, Eq. (32) is not even in agreement with the results of Ref. [26] related to the SF of classical plasma. This problem is caused during extension of dielectric function of quantum plasma which is the characteristic parameter of longitudinal potential space-charge waves and cannot be correct for the transverse electromagnetic wave description. In other theoretical study [52], for considering effect of electron pressure in zero temperature limit, being inspired by the longitudinal dielectric function of Eq. (26) and extending it to the relativistic regime by changing $m_0$ to $\gamma m_0$, the following equation has been employed for the dielectric function of thermal quantum plasma in the relativistic regime

$$\varepsilon = 1 - \frac{\omega_p^2}{\gamma \omega^2}\left(1 - \frac{\delta}{\gamma} - \beta'\right)^{-1}, \tag{33}$$

where $\beta' = k_0^2 V_F^2 / \omega_0^2$. Taking $\varepsilon_0 = 1 - \omega_p^2/\omega^2$, they obtained nonlinear part of the permittivity as follows

$$\Phi = \frac{\omega_p^2}{\omega_0^2}\left[1 - \frac{1}{\gamma}\left(1 - \frac{\delta}{\gamma} - \beta'\right)\right], \tag{34}$$

which after expanding the relativistic factor $\gamma$ with respect to the squared normalized laser beam amplitude, it includes a linear term of $-(\delta+\beta')\omega_p^2/\omega^2$ which is in contrast with the definition of $\Phi$ as the nonlinear part of the refractive index and would be considered in the linear permittivity $\varepsilon_0$. Furthermore, Eq. (34) does not confirm the results of Ref. [26] for the relativistic classical plasma after vanishing quantum effects by $\beta',\delta \to 0$. Equation (34), under any circumstances cannot be in accordance with Eq. (30) obtained here by considering the dynamics of electrons. In other quantum plasma SF theoretical studies [53-57], for some generalized physical conditions, e.g. magnetization of plasma or considering variable density for electrons, by the same approach of Refs. [50] and [52] the same problems are repeated during extension of the dielectric function to the relativistic case only by changing $m_0$ to $\gamma m_0$. It is worth mentioning that even if an appropriate dielectric function is chosen, changing $m_0$ to $\gamma m_0$ in order to modify a non-relativistic case to relativistic case without solving relativistic motion equation is equivalent to discard some terms such as $m_0 \mathbf{v} d\gamma/dt$ in the relativistic motion equation which can be the source of some incorrect terms of mentioned studies.

## V. Numerical discussions

Now, we are ready to study the evolution of laser spot size as a function of physical parameters numerically. In all investigated cases, we set normalized laser beam amplitude as $a_0^2=0.3$ and the beam waist $r_0=20\mu m$. As it is evident from Eq. (24), critical power for occurrence of SF is proportional to the laser beam intensity through normalized parameter $a_0^2$. Parameters in Figs. 1 and 2, have been chosen so that the laser power becomes greater than the critical power for stablishing the SF. The laser beam wave number in the medium, i.e. $k_0$, is determined by the linear dispersion relation (22).

Figure 1 shows the variations of the normalized laser spot size $r_s/r_0$ with respect to the normalized propagation distance $z/Z_R$ for two different electron densities: $n_0=4\times 10^{19} cm^{-3}$ and $n_0=4\times 10^{20} cm^{-3}$ while the laser frequency is $\omega_0=5\times 10^{20} s^{-1}$. SF of quantum plasma has been plotted with solid line while the dashed line denotes the classical plasma in which we have discarded terms related to the quantum effects in the denominator of Eq. (20). For both cases of different densities, laser beam catastrophically focuses in a propagation length less than the Rayleigh length. We can see that an increase in the electron density improves the SF property of the plasma. For both cases, quantum effects lead to a decrease in the nonlinearity of plasma and consequently to the weakening the SF. As one can find from Eq. (20), two terms containing quantum label $\hbar$ in the denominator, i.e. $\hbar^2 k_0^4/(m^2\omega_0^2)$ and $n_0\beta k_0^2/\omega_0^2$, are related to the quantum effects and their existence leads to the reduction of nonlinear term $N$. Numerical calculations show that the Bohm term: $\hbar^2 k_0^4/(m^2\omega_0^2)$, is effectively dominant. As an example, for the case $\omega_0 = 1.8\times 10^{20} s^{-1}$ and $n_0 = 4\times 10^{20} cm^{-3}$ the values of Bohm and degenerate gas pressure terms are $\hbar^2 k_0^4/(m^2\omega_0^2) \approx 0.053$ and $n_0\beta k_0^2/\omega_0^2 \approx 10^{-8}$. For higher frequencies or lower wavelengths from the linear dispersion relation (22) one can consider the approximate relation $\omega_0 \approx k_0 c$ which implies that the Bohm term approximately is proportion to the $\omega_0^2$ thus decreasing the laser beam frequency can lead to a decrease in the Bohm term. Additionally, to find a better sense about the occurrence of quantum limits, let us calculate some characteristic parameters of the system. For the case $\omega_0 = 1.8\times 10^{20} s^{-1}$ and $n_0 = 4\times 10^{20} cm^{-3}$ the laser beam wavelength inside the plasma is $\lambda_0 \approx 0.1 A^0$ and the average separation of electrons is $d = n_0^{-1/3} \approx 0.13 A^0$. For the quantum plasmas these lengths should be in the order or smaller than the de Broglie wavelength and Fermi length, which are $\lambda_B = \dfrac{\hbar}{\sqrt{2\pi m k_B T}} \approx 7.15 A^0$ and $\lambda_F = \dfrac{V_F}{\omega_p} \approx 1.34 A^0$, respectively, that implies the existence of quantum limit.

In the figure (2), we have decreased the laser beam frequency in comparison with the previous case and we set it as $\omega_0 = 1.8\times 10^{20} s^{-1}$. As mentioned before, decreasing the frequency, causes a decrease in quantum effects and consequently leads to the SF quality improvement. It is interesting to note that, for both selected values for the electron density, quantum effects on the SF are negligible.

**VI. Conclusions**

In this paper, we investigated the SF of an intense X-ray laser beam propagating through a quantum plasma. After obtaining a nonlinear wave equation, the spot size evolution of the laser beam was studied. We have considered nonlinear relativistic dynamics of degenerate electrons via solving momentum equation by means of a perturbative method. Results of our method were compared with the methods in which extended longitudinal permittivity of quantum plasma is used for describing the nonlinear dynamics of degenerate electrons. It is shown that our approach is more reliable, and its results are acceptable. Furthermore, it was found that considering the quantum effects, leads to the reduction of nonlinearity behavior of system and consequently to the decrease in the focusing property of the medium. Numerical studies have shown that the dominant part of quantum effects comes from the Bohm potential term. Additionally, it was shown that, for the dens and cold plasmas, the quantum effects will be noticeable only for the sufficiently short wavelengths, because the Bohm potential term is proportional to the squared frequency of the laser beam.

**Acknowledgment**

This work was partially supported by the Ferdowsi University of Mashhad under Grant No. 3/43953.

**Figures Caption**

Fig. 1. The variations of the normalized laser spot size $r_s/r_0$ with respect to the normalized propagation distance $z/Z_R$ for two different electron density of a) $n_0 = 4\times10^{19} cm^{-3}$ ,b) $n_0 = 4\times10^{20} cm^{-3}$ when the laser frequency is $\omega_0 = 5\times10^{20} s^{-1}$. Dashed curves show the spot size evolution without considering quantum effects.

Fig. 2. The variations of the normalized laser spot size $r_s/r_0$ with respect to the normalized propagation distance $z/Z_R$ for two different electron density of a) $n_0 = 4\times10^{19} cm^{-3}$ ,b) $n_0 = 4\times10^{20} cm^{-3}$ when the laser frequency is $\omega_0 = 1.8\times10^{20} s^{-1}$. Dashed curves show the spot size evolution without considering quantum effects.

a)

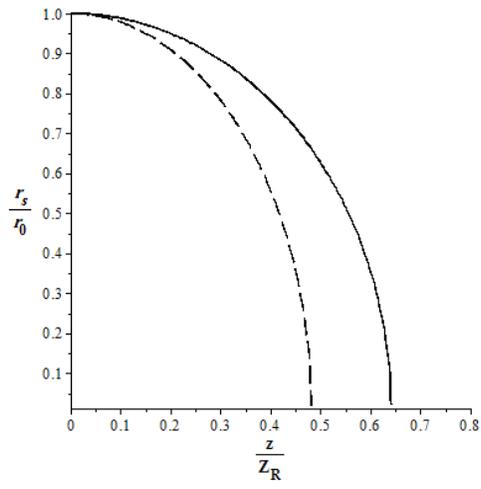

b)

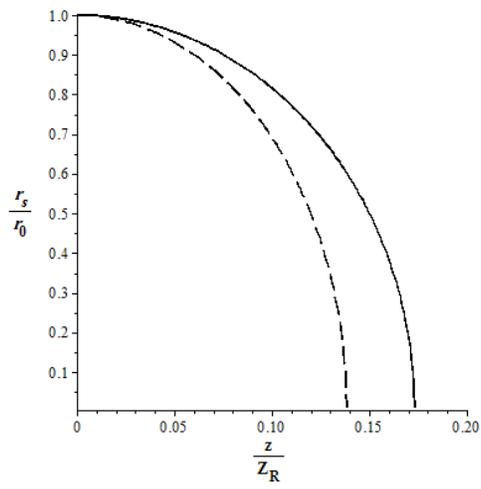

**Fig. 1**

a)

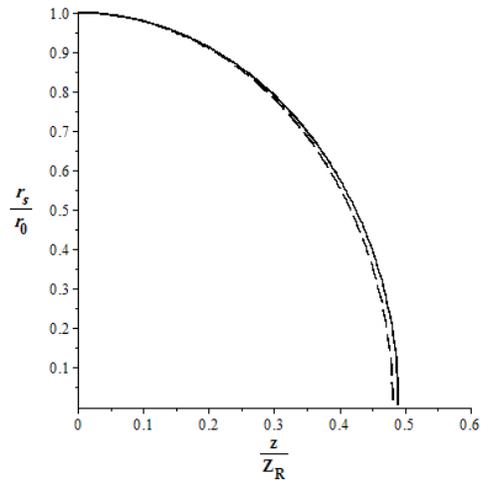

b)

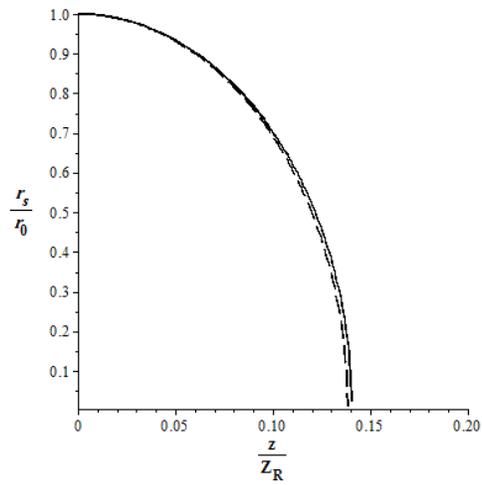

**Fig. 2**